\documentclass[cqg,12pt]{iopart}
\usepackage[centering,hmargin=2cm,vmargin=2.5cm]{geometry}
\usepackage[utf8]{inputenc}
\usepackage{graphicx}
\usepackage{amssymb}
\usepackage{amsfonts}
\usepackage{soul}
\usepackage{dsfont}
\usepackage{hyperref}
\usepackage{iopams}

\newcommand{\bra}[1]{\langle#1 \vert}
\newcommand{\ket}[1]{ \vert #1\rangle}
\newcommand{\braket}[2]{\langle#1 \vert #2\rangle}

\newcommand{\avg}[1]{\langle #1 \rangle}
\begin{document}
\date{\today}
\title[Quantum thermodynamics for a model of an expanding universe]{Quantum thermodynamics for a model of an expanding universe}

\author{Nana Liu$^1$, John Goold$^2$, Ivette Fuentes$^3$,Vlatko Vedral$^{1,4,5,6}$, Kavan Modi$^7$, David Edward Bruschi$^8$}

\address{$^1$ Clarendon Laboratory, Department of Physics, University of Oxford, Oxford OX1 3PU, United Kingdom}
\address{$^2$ The Abdus Salam International Centre for Theoretical Physics (ICTP), Trieste, Italy}
\address{$^3$ School of Mathematical Sciences, University of Nottingham, University Park, Nottingham NG7 2RD, United Kingdom}
\address{$^4$ Centre for Quantum Technologies, National University of Singapore, 3 Science Drive 2, Singapore 117543}
\address{$^5$ Department of Physics, National University of Singapore, 3 Science Drive 2, Singapore 117543}
\address{$^6$ Center for Quantum Information, Institute for Interdisciplinary Information Sciences, Tsinghua University, Beijing, 100084, China }
\address{$^7$ School of Physics and Astronomy, Monash University, Victoria 3800, Australia}
\address{$^8$ Racah Institute of Physics and Quantum Information Science Centre, the Hebrew University of Jerusalem, 91904 Givat Ram, Jerusalem, Israel}
\ead{Nana.Liu@physics.ox.ac.uk}

\begin{abstract}
We investigate the thermodynamical properties of quantum fields in curved spacetime. Our approach is to consider quantum fields in curved spacetime as a quantum system undergoing an out-of-equilibrium transformation. The non-equilibrium features are studied by using a formalism which has been developed to derive fluctuation relations and emergent irreversible features beyond the linear response regime. We apply these ideas to an expanding universe scenario, therefore avoiding assumptions on the relation between entropy and quantum matter. We provide a fluctuation theorem which allows us to understand particle production due to the expansion of the universe as an entropic increase. Our results pave the way towards a different understanding of the thermodynamics of relativistic and quantum systems in our universe.
\end{abstract}

\maketitle

\tableofcontents
%---------------------------------------------------------------------------------------------------------------------------------------------------------------------------------------------%
\section{Introduction}
%---------------------------------------------------------------------------------------------------------------------------------------------------------------------------------------------%

The study of our universe is an old and ongoing challenge and has allowed us to understand many physical phenomena and unveil new puzzles. Among the most interesting open questions in modern research are the origin of the accelerating expansion of spacetime \cite{1538-3881-116-3-1009, 0004-637X-517-2-565} and the source of the current entropy content of the universe \cite{Guth:81,0264-9381-26-14-145005, 0004-637X-710-2-1825}. To answer these questions requires applying thermodynamics to cosmology. Thermodynamics is one of the most exportable branches of physics and has successfully been applied to understand small and large systems, including cosmological models. It has provided important predictions for entropy-matter relations in expanding cosmological models \cite{Guth:81,Frautschi13081982} and entropy bounds for black hole scenarios \cite{Bekenstein:72, PhysRevD.7.2333, PhysRevD.9.3292}.

Matter contributes to the large entropy content of the universe \cite{Guth:81, 0264-9381-26-14-145005,  0004-637X-710-2-1825}. The emergence of the matter (starting from the vacuum) could be a consequence of the expansion of the spacetime \cite{PhysRevLett.21.562,  PhysRev.183.1057,  PhysRevD.3.346}. Therefore it is often assumed, but without proof, that entropy production should be directly related to particle creation \cite{prigogine}. However, the laws of physics are fully reversible which has led to the conclusion that entropy cannot be increased in processes governed by physical laws such as Einstein and Schr\"odinger equations \cite{prigogine}. This apparent contradiction remains an unsolved puzzle today and has led to a body of work aimed at understanding the role of particle creation in the thermodynamical processes of our universe.

In the past decades there have been quantitative approaches to the question of cosmological entropy production that employ both open and closed, classical and quantum formalisms \cite{prigogine, hureview, hu1986, hulin,  hukandrup}. Often in these contexts (von Neumann) entropy is viewed as quantifying inaccessible information or coarse-grained information (by necessity or choice). However, the von Neumann entropy of closed quantum systems cannot change under unitary evolution. This sparked approaches based on open classical and quantum systems \cite{prigogine, hureview}. Quantum approaches allow for the state of matter to become more mixed as the system interacts with its environment. Consequently, one witnesses a loss of knowledge or information about the system and increase of entropy. It is argued that entropy production is directly related to the number of particles created \cite{kokshu}. Other work suggests that in closed quantum systems one might still expect to find an entropic quantity that increases directly as a consequence of the creation of particles due to expansion \cite{hu1986}. However, a firm connection between an established measure of entropy and the amount of particles created in cosmological scenarios remains an open question.

In this work we jointly use tools from quantum field theory in curved spacetime \cite{BandD} and the recently developed concepts from thermodynamics of quantum systems \cite{campisireview} to investigate a relationship between entropy production and particle creation in an expanding universe. This quantum thermodynamic approach has been recently applied to study dissipation and entropy production in a variety of models in many-body physics \cite{silva,  ross1,  dudu,  splitting,  sindona}. We explore applications of this approach to a simple model for cosmological expansion. We give a thermodynamic meaning to particle creation in terms of a quantity called \emph{inner friction} \cite{kosloff,  engine,plastina}. We show that inner friction arises due to the quantum fluctuations of the fields and has an entropic interpretation stemming from a quantum fluctuation relation \cite{campisireview}. Our main result is a quantum version of the second law of thermodynamics for an expanding universe which accounts for the creation of matter. 
The question we are addressing here differs from those considered in work connected to inflation \cite{Guth:81}, since we are considering whether or not there is a purely quantum mechanical contribution to entropy associated with the unitary process of quantum particle creation.
The techniques developed in this work can be used to better understand cosmological processes or scenarios based on similar mathematical descriptions.

%---------------------------------------------------------------------------------------------------------------------------------------------------------------------------------------------%
\section{Quantum Field Theory in Curved Spacetime}
%---------------------------------------------------------------------------------------------------------------------------------------------------------------------------------------------%

We consider particles arising from excitations of quantum fields that propagate on a classical spacetime (the semi-classical regime). We will use the natural units convention $c=\hbar=G=k_B=1$. We consider for simplicity a massive scalar quantum field $\phi(x,t)$ with mass $m$ in (3+1)-dimensional spacetime \cite{BandD} with metric $g_{\mu \nu}$. An example of a massive scalar field in cosmology could be the inflation field \cite{mukhanov}. 
The equation of motion of the field, which is the Klein-Gordon equation in curved spacetime, can be written in the form
\begin{eqnarray}
(\square+m^2)\phi=0,
\end{eqnarray}
where the d'Alembertian is $\square \equiv (\sqrt{-g})^{-1} \partial_{\mu} [\sqrt{-g} g^{\mu \nu} \partial_{\nu}]$.

The field can be decomposed in any orthonormal basis of solutions $u_{\boldsymbol{k}} (t,\boldsymbol{x})$ to the Klein-Gordon equation as 
\begin{eqnarray}
\phi  =\int d^3\,\boldsymbol{k}\,\left[a_{\boldsymbol{k}}\,u_{\boldsymbol{k}} + a^{\dagger}_{\boldsymbol{k}}\,u^*_{\boldsymbol{k}}\right],
\end{eqnarray}
with annihilation and creation operators $a_{\boldsymbol{k}}$, $a^{\dagger}_{\boldsymbol{k}}$ that satisfy the canonical commutation relations $[a_{{\boldsymbol{k}}'},a^{\dagger}_{\boldsymbol{k}}]=\delta^3({\boldsymbol{k}}-{\boldsymbol{k}}')$ and all others vanish. The annihilation operators $a_{\boldsymbol{k}}$ define the vacuum state $\ket{0}$ through $a_{\boldsymbol{k}} \ket{0}=0, \,\forall {\boldsymbol{k}}$. In general, it is convenient to choose the set of modes $\{u_{\boldsymbol{k}}\}$ if it satisfies (at least asymptotically) an eigenvalue equation of the form $i\,\partial_\tau u_{\boldsymbol{k}}=\omega_{\boldsymbol{k}}\,u_{\boldsymbol{k}}$, where $\partial_\tau$ is some (possibly global) time-like Killing vector and $\omega_{\boldsymbol{k}}$ is a real eigenvalue. 

However, a different, but still convenient, choice of modes $\{ \tilde{u}_{\boldsymbol{k}}(x, t) \}$ might exist which satisfy the same equation of motion. The field can be expanded in terms of these modes as $\phi  =\int \,d^3{\boldsymbol{k}}\,\left[\tilde{a}_{\boldsymbol{k}}\, \tilde{u}_{\boldsymbol{k}} + \tilde{a}^{\dagger}_{\boldsymbol{k}}\, \tilde{u}^*_{\boldsymbol{k}}\right]$, where $\tilde{a}_{\boldsymbol{k}}$, $\tilde{a}_{\boldsymbol{k}}^{\dagger}$ are new ladder operators for these modes that satisfy commutation relations $[\tilde{a}_{{\boldsymbol{k}}'},\tilde{a}^{\dagger}_{\boldsymbol{k}}]=\delta^3({\boldsymbol{k}}-{\boldsymbol{k}}')$ while all other commutators vanish. These operators define a different vacuum $\ket{\tilde{0}}$ through $\tilde{a}_{\boldsymbol{k}'} \ket{\tilde{0}}=0$. If one is fortunate, one is able to find such an alternative candidate for the vacuum state. It may be different to $\ket{0}$, but is still equivalent to it (i.e., residing in the same Hilbert space) \footnote{However, if there exist inequivalent vacua, such as those those residing in different Hilbert spaces, the predictions for different observers in curved spacetime cannot be compared. Although this is a common problem in quantum field theory in curved spacetime, this is not the case for the examples discussed in this paper.}. This feature of having more than one, but equivalent, vacuum is a central property of quantum field theory in curved spacetime and it is at the core of some of the most exciting predictions of this theory, such as black hole evaporation \cite{Hawking:74}, the Unruh effect \cite{unruh} and the dynamical Casimir effect \cite{Dodonov:10}. 

The two sets of creation and annihilation operators are related to each other by a Bogoliubov transformation 
\begin{eqnarray}
\tilde{a}_{\boldsymbol{k}}=\int_{{\boldsymbol{k}}'} d{\boldsymbol{k}}'\left[ \alpha_{{\boldsymbol{k}}{\boldsymbol{k}}'} a_{{\boldsymbol{k}}'}+\beta^*_{{\boldsymbol{k}}{\boldsymbol{k}}'} a_{{\boldsymbol{k}}'}^{\dagger}\right],
\end{eqnarray}
where $\alpha_{{\boldsymbol{k}}{\boldsymbol{k}}'}$, $\beta_{{\boldsymbol{k}}{\boldsymbol{k}}'}$ are the well-known Bogoliubov coefficients. They satisfy the Bogoliubov identities (see \cite{BandD}) and encode information about the spacetime (i.e., expansion rate of the expanding universe \cite{ivetteentanglement} or the mass of a black hole \cite{Hawking:74}). Particle production in quantum field theory scenarios is a direct consequence of the difference between vacua \cite{BandD} residing in the same Hilbert space. For example, it was shown that the initial vacuum state $\ket{0}$ \textit{evolves} into an excited state in the expanding universe or black hole scenarios \cite{BandD} or that a uniformly accelerated observer perceives the inertial vacuum $\ket{0}$ as a thermal bath of particles \cite{unruh}. In all of these cases, the content of particles is $\avg{n_{out}^{\boldsymbol{k}}} =\bra{0} \tilde{a}_{\boldsymbol{k}}^{\dagger} \tilde{a}_{\boldsymbol{k}}\ket{0}=\int d^3l\,|\beta_{{\boldsymbol{l}}{\boldsymbol{k}}}|^2 $, which underlines the role of the Bogoliubov transformations.

%---------------------------------------------------------------------------------------------------------------------------------------------------------------------------------------------%
\subsection{Two-mode squeezing and cosmology\label{sec:twomodecosmo}}
%---------------------------------------------------------------------------------------------------------------------------------------------------------------------------------------------%

We now specialise to the Robertson-Walker spacetime in $1+1$ dimensions with coordinates $(t,x)$. The line element is $ds^2=-dt^2+a^2(t)dx^2=\Omega^2(\eta)(-d \eta^2+dx^2)$, where $a(t)$ is the scale factor and $\Omega^2(\eta)$ is the conformal scale factor. The conformal time $\eta$ is defined by $d \eta=dt/a(\eta)$. We note that considering $1+1$ dimensions introduces significant technical and conceptual simplifications, which allow us to obtain analytical results which elucidate the connection between particle creation and irreversibility in the dynamics of spacetime. We leave it to future work to extend our results to more realistic models which include all $3+1$ dimensions.

We notice that the (1+1)-dimensional Klein-Gordon equation, in the convenient coordinates $\eta,x$, reduces to a Klein-Gordon equation in Minkowski spacetime
\begin{eqnarray}
\left(-\frac{\partial^2}{\partial\eta^2}+\frac{\partial^2}{\partial x^2}+m^2\Omega^2(\eta)\right)\phi(x,\eta)=0.
\end{eqnarray}
There are two plane wave solutions to the field equation, the ``in'' modes $u_k=(1/\sqrt{4 \pi \omega})\exp(ikx-i\omega \eta)$ in the asymptotic past and the ``out'' modes $\tilde{u}_k=(1/\sqrt{4 \pi \tilde{\omega}})\exp(ikx-i\tilde{\omega}\eta)$ in the asymptotic future, with frequencies
\begin{eqnarray} 
\omega_k &=\sqrt{k^2+m^2 \Omega^2\vert_{\eta \rightarrow -\infty}}, \nonumber \\
 \tilde{\omega}_k &=\sqrt{k^2+m^2 \Omega^2\vert_{\eta \rightarrow +\infty}}.
\end{eqnarray}

Isotropy, conservation of momentum and energy \cite{BandD} simplify the Bogoliubov transformation between the ``in'' and ``out'' bosonic operators to 
\begin{eqnarray} \label{eq:atildea}
\tilde{a}_k=\alpha_k a_k+\beta^*_k a^{\dagger}_{-k},
\end{eqnarray}
known in quantum optics as two-mode squeezing \footnote{The generation of two-mode squeezed states is well-established in quantum optics and further references can be found in \cite{gerryknight}.}. This generates an entangled state with strong correlations between modes $k$ and $-k$ of the field. This operation creates or annihilates particles pair-wise in the $(k,-k)$ mode pair, thus can be likened to the creation or annihilation of particle/anti-particle pairs from the vacuum. The connection between two-mode squeezing operations and quantum field theory has already proved useful in approaching the topic of particle creation in cosmology \cite{grishchuk,  hu1994, ivetteentanglement, Fuentes:Mann:10}, the Unruh effect \cite{PhysRevA.76.062112} and Hawking radiation \cite{adesso:correlation}. Two-mode squeezing is a generic feature in some of the key predictions of quantum field theory in curved spacetime. The Bogoliubov coefficients in Eq.~\ref{eq:atildea} satisfy $|\alpha_k|=\cosh(z_k)$, $|\beta_k|=\sinh(z_k)$, where $z_k$ is known as the squeezing parameter \cite{barnett}. We choose the conformal form factor  $\Omega^2(\eta) = 1 + \epsilon ( 1 + \tanh(\sigma\,\eta))$, where $\epsilon,\sigma>0$ govern the total volume and rate of expansion respectively and we have $\tanh^2 (z_k)=\sinh^2 (\pi (\tilde{\omega}-\omega)/2 \sigma)/(\sinh^2 (\pi (\tilde{\omega}+\omega)/2 \sigma))$.

In this work we use the Heisenberg picture. The initial (asymptotic past) and the final (asymptotic future) Hamiltonians are
\begin{eqnarray} \label{eq:hamneginfty}
\mathcal{H} (x, \eta \rightarrow -\infty) &=\int_k dk\frac{\omega_k}{2}[a_k^{\dagger} a_k+a_k a_k^{\dagger}], \nonumber \\
 \mathcal{H} (x, \eta \rightarrow \infty) &=\int_k dk\frac{\tilde{\omega}_k}{2}[\tilde{a}_k^{\dagger} \tilde{a}_k+\tilde{a}_k \tilde{a}_k^{\dagger}].
\end{eqnarray}
Eqs.~\ref{eq:atildea} and \ref{eq:hamneginfty} imply that time evolution for each pair of modes $(k,-k)$ is unitary. This means that there is interaction only between modes $k$ and $-k$ and no other modes of the field. Since each mode pair evolves independently under a change in spacetime, for simplicity we focus on one pair of modes $(k,-k)$ in order to illustrate our techniques. Thus we suppress all $k$ indices for the rest of this paper. We define $a_k \equiv a$ and $a_{-k} \equiv b$ throughout the rest of the work. The initial and final Hamiltonian are
\begin{eqnarray} \label{eq:hamiltonian}
H &=\omega (a^{\dagger} a+b^{\dagger}b+1), \nonumber \\
\tilde{H} &= \tilde{\omega}(\tilde{a}^{\dagger} \tilde{a}+\tilde{b}^{\dagger} \tilde{b}+1).
\end{eqnarray}
The dynamics of the quantum field is therefore determined uniquely by the initial and final mode frequencies $\omega, \tilde{\omega}$ and the squeezing parameter $z$. Here we note that the energy spectrum is equally spaced for both the initial and final Hamiltonian. 

We remark here that while in this case the evolution of the field may be considered to evolve unitarily between $H$ and $\tilde{H}$ during spacetime expansion \cite{cortez2015}, in more general scenarios, like in general (3+1)-dimensional cosmologies, defining unitary evolution is not straightforward and is sometimes not possible. However, in recent years there have been advances made in this area to show it is in fact possible to also get unitary evolution in cosmological $(3+1)$-dimensional models, see the recent progress in \cite{cortez2015, agullo2015} and references therein. 

%---------------------------------------------------------------------------------------------------------------------------------------------------------------------------------------------%
\section{Thermodynamics of Cosmology}
%---------------------------------------------------------------------------------------------------------------------------------------------------------------------------------------------%

Let us begin by specifying our main assumptions. (i) We can confine our attention to two modes of opposite momenta $(k, -k)$. This allows us to avoid difficulties with infinities that arise from using modes with sharp frequencies. Alternatively, we can extend our model to scenarios where particles are described by wavepackets instead and grey body factors are required. We leave this to future work. (ii) The ``speed'' and ``strength'' of interaction between the spacetime and the quantum field is much greater than the ``speed'' and ``strength'' of interaction between this field and any other external fields. In other words, we assume that any interaction between our field and others that might be present in the universe is negligible during the time it takes for the mode pairs of our field of interest to be correlated via two-mode squeezing. An example is in the inflationary scenario, where spacetime undergoes a very rapid expansion. Therefore, each mode pair evolves unitarily and is described by the Hamiltonian changing as $H \rightarrow \tilde{H}$. (iii) Our mode pair is approximated to start in a thermal state $\rho$, which is a subsystem of the universe. This comes from the result that any random subsystem will be in a thermal state for almost all pure states of the universe \cite{popescu}. This is due to interaction between our state and other fields, \textit{before} the spacetime begins expanding. 

Under these assumptions, we take advantage of the formal mathematical apparatus recently developed to study the thermodynamics of quantum systems driven out of equilibrium by unitary evolution \cite{kurchan,  tasaki,  talknerfluctuation,  Deffner1, campisireview} and we re-examine cosmological particle creation. We separate our thermodynamical framework into two components: a system (i.e., the pair of field modes) and a work reservoir (i.e., the spacetime). We view spacetime as the source for the change in the internal energy of the quantum field \cite{prigogine}. The action of the spacetime on the quantum field is treated as a purely classical work reservoir, in the same way that an external magnetic field or laser act as an energy source for a quantum optical system. Under this action, modes of opposite momenta undergo a unitary evolution as spacetime expands. The following results will apply to every pair of modes of the field.

%---------------------------------------------------------------------------------------------------------------------------------------------------------------------------------------------%
\subsection{Work done by spacetime}
%---------------------------------------------------------------------------------------------------------------------------------------------------------------------------------------------%

Since we view the classical spacetime as the external driving that takes our field away from equilibrium, there will be an energy change in the quantum field induced by the change in spacetime. In this sense, the changing spacetime is `doing work' onto the quantum field. This change of energy in the field will be called the average work. This is computed as the difference between the average energy corresponding to the final Hamiltonian $\tilde{H}$ compared to the initial Hamiltonian $H$ \cite{campisireview} as the field is unitarily evolved from $H$ to $\tilde{H}$. In this paper we continue to use the Heisenberg picture unless otherwise mentioned. A straightforward calculation gives the average work in our cosmological model:
\begin{eqnarray} 
\avg{W} & \equiv &\,\tr(\tilde{H}\, \rho)-\tr(H\, \rho) \nonumber\\
& = &\, \tilde{\omega}\, \avg{n_{cr}}+(\tilde{\omega}-\omega) (\avg{n_i} + 1),\label{eq:worknumber}
\end{eqnarray}
where $\rho$ is the state of our mode pair, $\avg{n_i}$ is the initial average number of excitations, and $\avg{n_{cr}}$ is the average number of created particles.

We identify three different contributions to work. The first term $\tilde{\omega} \avg{n_{cr}}$ is the work cost associated with the creation of new particles. The second term $(\tilde{\omega}-\omega) \avg{n_i}$ is the work cost in changing the frequencies of the particles already present in the initial thermal state. Finally, the cost $\tilde{\omega}-\omega$ of changing the ground state energy of the system. Note that the particle creation term $\tilde{\omega} \avg{n_c}$ does not arise from particle interaction, like particle decay and collisions (entropies in these other regimes are treated elsewhere, see \cite{hucosmology}).

If the expansion occurs in a quantum adiabatic limit \footnote{Note the distinction between thermal and quantum adiabaticity. Thermal adiabatic means that there is no heat, which is guaranteed for \emph{any} unitary dynamics. Quantum adiabatic means that there are no transitions between different energy levels during the evolution.}, the Bogoliubov coefficients $\beta_k$ vanish and the final adiabatic Hamiltonian is
\begin{eqnarray}\label{eq:adHam}
\tilde{H}_{ad} = \tilde{\omega} (a^{\dagger} a+b^{\dagger}b+1) = \frac{\tilde{\omega}}{\omega} H.
\end{eqnarray}
In this quantum adiabatic scenario, the average work done by spacetime onto the fields is defined as the adiabatic work $\avg{W}_{ad}$ which reads
\begin{eqnarray} \label{eq:adwork}
\avg{W}_{ad}= (\tilde{\omega}-\omega)(\avg{n_i}+1).
\end{eqnarray}
Note that no particles are created in an adiabatic evolution. This happens when either the rate of spacetime expansion is quasistatic (i.e., $\sigma \rightarrow 0$) or when the coupling between the field and spacetime disappears, which occurs for a massless scalar field. The difference between the average work $\avg{W}$ and the average adiabatic work $\avg{W}_{ad}$ defines the quantity $\avg{W}_{fric}$ called \textit{inner friction} \cite{kosloff,  engine, plastina}. In our cosmological setting the inner friction is directly proportional to particle creation
\begin{eqnarray}\label{eq:workfriction}
\avg{W}_{fric}:=\avg{W}-\avg{W}_{ad}= \tilde{\omega}\, \avg{n_{cr}}.
\end{eqnarray}
This result fits one's intuition that the more particles are created, the farther one is from a quasi-static evolution. When no particles are created (when the universe expands quasi-statically), there is still a work cost in expanding without inner friction being produced, which is quantified by $\avg{W}_{ad}$. Our final step is to show how inner friction $\avg{W}_{fric}$ can be interpreted as an entropy production, to be defined below, in the cosmological context.

%---------------------------------------------------------------------------------------------------------------------------------------------------------------------------------------------%
\subsection{Entropy Production and Cosmological Particle Creation}
%---------------------------------------------------------------------------------------------------------------------------------------------------------------------------------------------%
Inner friction can also be considered as quantifying entropy production during cosmological particle creation from the viewpoint of the entropy production fluctuation theorems. The fluctuation theorems were introduced first in classical systems by Evans \cite{evans} and Crooks \cite{Crooks} to define entropy production $s$ for systems when perturbed arbitrarily away from equilibrium and were later extended to quantum systems \cite{campisireview,tasaki}. These relations take the general form $e^s=P_F(s)/P_R(-s)$, where $P_F(s)$ is the probability distribution of $s$ when beginning from equilibrium. This is also called the ``forward'' distribution. $P_R(-s)$ is the (``reverse'') probability distribution of $s$ when a time-reversed driving is applied to the system starting at equilibrium. Thus entropy production, defined in this sense, expresses the difference between the forward and reverse probability distributions. 

This way of viewing entropy production motivates us to define ``forward'' and ``reverse'' processes in cosmology and then to derive a corresponding fluctuation theorem. We define our ``forward'' process to be the expansion of spacetime beginning in an equilibriums state of $H$, where this state is $\rho=\sum_j \exp(-E_j/T) \ket{j}\bra{j}/Z$ and $Z=\sum_j \exp(-E_j/T)$ is the partition function. Here $\{E_j\}$, $\{\ket{j}\}$ are the energy eigenvalues and eigenstates of $H$ and $T$ can be considered as the temperature of the state. The ``reverse'' process is the contraction of this spacetime but beginning in the final adiabatic Hamiltonian $\tilde{H}_{ad}$. The state $\rho$ remains the same since we are working in the Heisenberg picture. Now let $p_{n}$ be the probability of $n$ particles being found initially in one run of spacetime expansion. Let $q_{m}$ be the probability that $m$ particles are initially found in the spacetime contraction process. We can associate the entropic quantities $-\log(p_{n})$ and $-\log(q_{m})$ to these probabilities. We can then define the difference of these entropic quantities as $s_{nm}=-\log(q_{m})+\log(p_{n})$ where $p_{n}=\bra{n}\rho \ket{n}= \exp(-E_{n}/T)/Z$ and $q_{m}=\bra{m}\rho \ket{m}= \exp(-E_{m}/T)/Z$. We can thus rewrite our new entropic random variable as
\begin{equation}\label{eq:snm}
s_{nm}=\frac{E_m}{T}-\frac{E_n}{T}=\frac{\tilde{E}_m}{\tilde{T}}-\frac{E_n}{T},
\end{equation}
where $\{\tilde{E}_j\}$ are the eigenvalues of $\tilde{H}$ and we define an effective temperature $\tilde{T} \equiv \tilde{E}_j T/E_j$.  In our model, $\tilde{T}$ is a constant since the energy spectrum of $H$ and $\tilde{H}$ are equally spaced. This is also equivalent to $\tilde{T}=(\tilde{\omega}/\omega)T$ in our model.

We are now ready to define the probability distribution for an entropic quantity $s$ in the expansion process as
\begin{equation}\label{eq:pe}
P_E(s)=\sum_{n, m}\delta(s-s_{nm}) p_{m|n}p_{n},
\end{equation}
where $p_{m|n}=|\braket{\tilde{m}}{n}|^2$ and $\{\ket{\tilde{n}}\}$ are the eigenvectors of $\tilde{H}$. The term $p_{m|n}$ is the transition probability in going from $n$-particles in the beginning of expansion to $m$-particles at the end of expansion. Similarly, for the corresponding contraction process we can define $P_C(-s)=\sum_{n, m}\delta(s-s_{nm}) q_{n|m}q_{m}$. Here $q_{n|m}$ is the transition probability of going from $m$ particles to $n$ particles during spacetime contraction. Note that the normalisation conditions for both probability distributions are obeyed $\int P_E(s)ds=1=\int P_C(-s) ds$. Using Eqs.~\ref{eq:snm}, ~\ref{eq:pe} and the thermal state $\rho$ we find $\langle e^{-s} \rangle\equiv \int e^{-s}P_E(s)ds=\sum_{nm}e^{-s_{nm}}\braket{\tilde{m}}{n}\bra{n}\rho \ket{n}\braket{n}{\tilde{m}}=1.$ Combined with the normalisation condition the reverse probability distribution we find $\int \exp(-s)P_E(s)ds=1=\int P_C(-s)ds$, thus for our entropic quantity $s$ we have the following fluctuation relation 
\begin{equation}
e^s=\frac{P_E(s)}{P_C(-s)}.
\end{equation}
This suggests the process in which $s$ is positive is exponentially more likely in the spacetime expansion case compared to the contraction process. Taking the logarithm on both sides and taking the average with $P_E(s)$ we have
\begin{eqnarray}
\avg{s} \equiv \tilde{T}\int s P_E(s)ds= K[P_E(s) \| P_C(-s)], \label{entpos}
\end{eqnarray}
where $K[P_E(s) \| P_C(-s)]$ is the Kullback-Leibler divergence (or relative entropy) between probability distributions $P_E(s)$ and $P_C(-s)$ \cite{kullbackleibler}. This entropic quantity $\avg{s}$ is positive since the relative entropy $K[X\|Y]$ is positive. It vanishes only when $P_E(s)=P_C(-s)$, i.e., for an adiabatic expansion of the spacetime, where no particles are created. This can be understood from the connection between this entropic quantity and particle creation. We show this relationship explicitly by demonstrating $\langle s \rangle$ is also proportional to our inner friction term. Using Eq.~\ref{eq:pe} and $\rho=\sum_j p_j \ket{j}\bra{j}$ we have
\begin{eqnarray}
\tilde{T} \langle s \rangle \equiv \int sP_E(s)ds=\tilde{T} \sum_{mn} s_{nm}\braket{\tilde{m}}{n}p_n\braket{n}{\tilde{m}}.
\end{eqnarray}
Then inserting Eqs.~\ref{eq:adHam}, ~\ref{eq:workfriction} and ~\ref{eq:snm} we derive 
\begin{eqnarray}\label{eq:sandW}
\tilde{T} \langle s \rangle
                                        &=\sum_{m} \tilde{E}_m\bra{\tilde{m}}\sum_n p_n \ket{n}\braket{n}{\tilde{m}}-\frac{\tilde{T}}{T} \sum_{n}E_n p_n \nonumber \\
                                        &=\tr(\tilde{H}\rho)-\frac{\tilde{\omega}}{\omega} \tr(H \rho)=\langle W_{fric} \rangle.
\end{eqnarray}
From Eqs.~\ref{eq:workfriction} and ~\ref{eq:sandW} we now have an exact relationship between an entropy production and the number of particles created
\begin{equation}\label{eq:snumber}
\langle s \rangle =\frac{\langle W_{fric}\rangle}{\tilde{T}}=\frac{\tilde{\omega}}{\tilde{T}} \langle n_{cr} \rangle.
\end{equation}
Since $\langle s \rangle \geq 0$, this implies that inner friction is also positive. The positivity of $\avg{W}_{fric}$ can be seen as a statement of the second law of thermodynamics \cite{jarzynski} in a statistical formulation \footnote{It has been shown that inner friction is non-negative for any time-dependent Hamiltonian starting from a passive state \cite{alla} and non-positive for an active state \cite{campisi2011}. A thermal state, which is the assumption used in this paper, is an example of a passive state.}. This is strong evidence that $\avg{s}$ should be considered a suitable entropic term (similar to the entropy production as originally defined by Crooks \cite{Crooks}) to use in this cosmological context. This is the main result of this paper. 

The intimate relationship between this particular measure of entropy production in spacetime expansion and particle creation is another main result in this paper. We observe that if a state diagonal in the number basis (e.g. thermal state) undergoes two-mode squeezing we find $\avg{n}_{cr} \geq 0$. From Eq.~\ref{eq:snumber} we see that this is not only consistent with the second law of thermodynamics $\avg{W}_{fric} \geq 0$ but it also provides an alternative interpretation for $\avg{n}_{cr} \geq 0$ in terms of the second law. 

Inner friction can also be related to the quantum relative entropy \cite{plastina} \footnote{This formulation in terms of the quantum relative entropy is more transparent in the Schrodinger picture representation.}
\begin{eqnarray}\label{fric:to:rel:ent}
\avg{W}_{fric} = \tilde{T} \, K[\rho_f || \rho_{ad}],
\end{eqnarray}
where $\rho_f$ and $\rho_{ad}$ are the final states in the Schrodinger picture after actual spacetime expansion and adiabatic expansion respectively. The positivity of the quantum relative entropy has already been related to the second law of thermodynamics \cite{sagawa}. Eqs.~\ref{entpos} and \ref{fric:to:rel:ent} relate cosmological particle creation directly to a classical and a quantum relative entropy. 
%---------------------------------------------------------------------------------------------------------------------------------------------------------------------------------------------%
\subsection{Extension to other scenarios}
%---------------------------------------------------------------------------------------------------------------------------------------------------------------------------------------------%

The techniques developed here apply in a straightforward fashion to any quantum field theoretical or physical scenario that involves two mode squeezing. Scenarios of this type include the well-known Unruh effect \cite{unruh} and to the radiating eternal massive black hole scenario \cite{BandD}. In the Unruh effect there is one stationary observer and another observer uniformly accelerated with respect to the first observer with uniform acceleration $a$. In the radiating black hole scenario, the interesting parameter is the black hole mass $M$. In both cases, the ``inertial'' vacuum is perceived by a stationary observer as a state full of particles which are thermally distributed with a temperature $T_U\propto a$ in the Unruh case and $T_H\propto 1/M$ in the black hole case. Furthermore, particles are produced in correlated pairs in the exact same fashion as described in this work, i.e., two-mode squeezing. The squeezing $z$ satisfies $\tanh z=\exp[-\omega/T_U]$ for the Unruh case and analogously $\tanh z=\exp[-\omega/T_H]$ for the black hole case. Given these relations between squeezing and parameters of the physical setups, one can immediately apply the techniques developed here to find how particle number (entropy production) is expressed in the Unruh effect and radiating black hole scenario. 

However, in the Unruh effect and in the eternal black hole scenario, there is an added complication that event horizons are present. This suggests that observers will be able to access only one mode of the field and therefore will experience non-unitary dynamics. We leave it to further work to apply techniques of open quantum systems to analyse the interpretation of these physical processes by localised observers \cite{Liu:Goold:future}.

%---------------------------------------------------------------------------------------------------------------------------------------------------------------------------------------------%
\section{Discussion}
%---------------------------------------------------------------------------------------------------------------------------------------------------------------------------------------------%
We start from a thermal state of a massive scalar field in an expanding classical (1+1)-dimensional Robertson-Walker spacetime. The momentum $k$ and $-k$ modes of the field interact unitarily during this spacetime expansion and can be modelled by two-mode squeezing. This scenario is likened to a quantum optical system undergoing two-mode squeezing that is driven by a classical external source. The classical spacetime background plays the role of this classical external driver for the quantum field and takes the field away from equilibrium as the spacetime expands. 

Since two-mode squeezing only introduces interactions between mode $k$ and its partner $-k$, there is unitary evolution for each such mode pair. This allows us to examine each mode pair independently and to consider our mode pairs and external spacetime together forming a closed system during spacetime expansion. To ensure this, we assume that any interaction between our field and any other field that might be present during spacetime expansion to be negligible. For example, this can occur during the inflationary scenario when spacetime undergoes very rapid expansion. We also show that the Hamiltonian of these mode pairs corresponding to the initial and final stage of expansion have an equally spaced energy spectrum, like a harmonic oscillator. We have also assumed an initial thermal state and in future work it would be interesting to study deviations from the thermal state.

With the above assumptions, we show how a quantity called inner friction is proportional to the number of particles created during spacetime expansion and is positive. This quantity is defined as the difference between the average work done by an expanding spacetime onto the field and the average work done by the spacetime if it had expanded adiabatically. Inner friction is also linked with an entropy production, defined in the sense of entropy production fluctuation theorems.  

In summary, we employed tools from quantum field theory and quantum thermodynamics to study the connection between entropy production and creation of particles in quantum field theoretical setups. Our main result was to provide a second law of thermodynamics for closed relativistic and quantum systems which explains how a unitary process, such as particle production, is connected to an increase of entropy. Our work is free from assumptions on the relation between matter and entropy and provides and intuitive understanding of how energy is used in the process of expansion and particle creation in a cosmological scenario. Furthermore, our formalism extends to all setups where the unitary evolution breaks down into a collection of two mode squeezing operations. Our work opens a new avenue for the understanding of the quantum thermodynamics of our universe. 

\ack
We thank M. Barbieri, J. D. Bekenstein, O. Dahlsten, B. L. Hu, M. Huber, and A. Lee, for comments and discussions, J. Thompson for a careful reading of the manuscript and two anonymous referees for their careful comments on the manuscript. N. L. extends a special thanks to B. L. Hu (``Hu Lao Shi") for introducing this topic to her and for his inspiration and encouragement. N. L. thanks Monash University and the Hebrew University of Jerusalem for hospitality during the completion of this work. N. L. was supported by the Clarendon Fund and Merton College of the University of Oxford. I. F. acknowledges support from EPSRC (CAF Grant No. EP/G00496X/2). This work was partially supported by the COST Action MP1209. V. V. acknowledges funding from the EPSRC, the Templeton Foundation, the Leverhulme Trust, the Oxford Martin School, the National Research Foundation (Singapore), the Ministry of Education (Singapore) and from the EU Collaborative Project TherMiQ (Grant Agreement 618074). D. E. B. was supported by the I-CORE Program of the Planning and Budgeting Committee and the Israel Science Foundation (grant No. 1937/12), as well as by the Israel Science Foundation personal grant No. 24/12.

%---------------------------------------------------------------------------------------------------------------------------------------------------------------------------------------------%
\section*{References}
%---------------------------------------------------------------------------------------------------------------------------------------------------------------------------------------------%

\bibliographystyle{iopart-num}
\bibliography{RefCosmopaper}

\end{document}